\documentclass[twocolumn,aps,prl,showpacs,superscriptaddress,floatfix]{revtex4}
\usepackage{graphicx}
\usepackage{dcolumn}
\usepackage{epsf}
\usepackage{bm}
\usepackage{times}
\newcommand\vb{\vbox to 9 pt{}}

\newcommand\T{\rule{0pt}{2.4ex}}
\newcommand\B{\rule[-0.8ex]{0pt}{0pt}}
\newcommand\ri{{\rm i}}
\newcommand\iG{{\it \Gamma}}

\usepackage{graphicx}

\newcommand{\addrGaithersburg}{National Institute of Standards and Technology, 
Gaithersburg, MD 20899-8420, USA}
\newcommand{\addrHeidelberg}{Max--Planck--Institut f\"ur Kernphysik,
Saupfercheckweg 1, 69117 Heidelberg, Germany}
\begin{document}

\title{Fundamental constants and tests of theory in Rydberg states of hydrogen-like ions}

\author{Ulrich D.~Jentschura}
\affiliation{\addrGaithersburg}
\affiliation{\addrHeidelberg}

\author{Peter J.~Mohr}
\affiliation{\addrGaithersburg}

\author{Joseph N.~Tan}
\affiliation{\addrGaithersburg}

\author{Benedikt J.~Wundt}
\affiliation{\addrHeidelberg}

\date{\today}

\begin{abstract}

Comparison of precision frequency measurements to quantum
electrodynamics (QED) predictions for Rydberg states of hydrogen-like
ions can yield information on values of fundamental constants and test
theory.  With the results of a calculation of a key QED contribution
reported here, the uncertainty in the theory of the energy levels is
reduced to a level where such a comparison can yield an improved value
of the Rydberg constant.

\end{abstract}
\maketitle
%

Quantum electrodynamics (QED) makes extremely accurate predictions
despite the ``mathematical inconsistencies and renormalized infinities
swept under the rug" \cite{dyson}.  With the assumption that the theory
is correct, it is used to determine values of the relevant fundamental
constants by adjusting their values to give the best agreement with
experiments \cite{2005191}.  In this paper, we consider the possibility
of making such comparisons of theory and experiment for Rydberg states
of cooled hydrogen-like ions using an optical frequency comb.  We find
that because of simplifications in the theory that occur for Rydberg
states, together with the results of a calculation reported here, the
uncertainty in the predictions of the energy levels is dominated by the
uncertainty in the Rydberg constant, the electron-nucleus mass ratio,
and the fine-structure constant.  Apart from these sources of
uncertainty, to the extent that the theory remains valid, the
predictions for the energy levels appear to have uncertainties as small
as parts in $10^{17}$ in the most favorable cases.

\vskip -1 pt

The CODATA recommended value of the Rydberg constant has been obtained
primarily by comparing theory and experiment for twenty-three transition
frequencies or pairs of frequencies in hydrogen and deuterium
\cite{2005191}.  The theoretical value for each transition is the
product of the Rydberg constant and a calculated factor based on QED
that also depends on other constants.  While the most accurately
measured transition frequency in hydrogen (1S--2S) has a relative
uncertainty of $1.4\times10^{-14}$ \cite{2005082}, the recommended value
of the Rydberg constant has a larger relative uncertainty of
$6.6\times10^{-12}$ which is essentially the uncertainty in the
theoretical factor.  The main source is the uncertainty in the charge
radius of the proton with additional uncertainty due to uncalculated or
partially calculated higher-order terms in the QED corrections.  This
uncertainty could be reduced by a measurement of the proton radius in
muonic hydrogen \cite{2007129}, or by a sufficiently accurate
measurement of a different transition in hydrogen.  On the other hand,
for Rydberg states, the fact that the wave function is small near the
nucleus results in the finite nuclear size correction being completely
negligible.  Also, for Rydberg states, the higher-order terms in the QED
corrections are relatively smaller than they are for S states, so
theoretical expressions with a given number of terms are more accurate.

\begin{figure}
\resizebox{!}{5.8 cm}{
\includegraphics{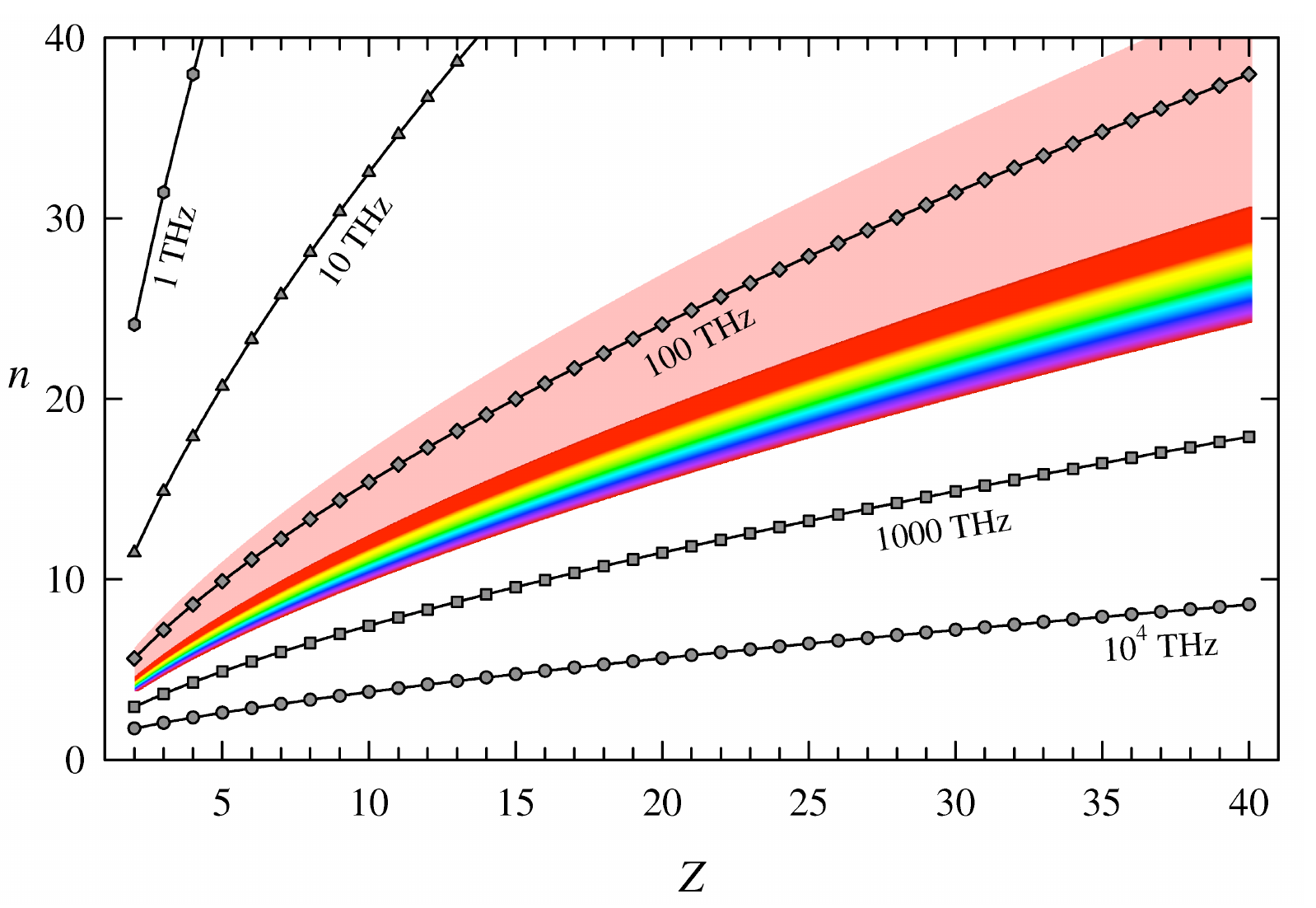}}
\caption{ \label{fig:nvsz}
Graph showing values of $Z$ and approximate $n$ that give a specified
value of the frequency for transitions between states with principal
quantum number $n$ and $n-1$ in a hydrogen-like ion with nuclear charge
$Z$.  Frequencies in the near infrared and visible range are indicated
in color.}
\end{figure}

Circular Rydberg states of hydrogen in an 80 K atomic beam have been
studied with high precision for transition wavelengths in the millimeter
region, providing a determination of the Rydberg constant with a
relative uncertainty of $2.1\times10^{-11}$ \cite{pc03dc,th01dv}.  With
the advent of optical frequency combs \cite{2006403}, precision
measurements of optical transitions between Rydberg states have now
become possible using femtosecond lasers.  An illustration is the laser
spectroscopy of antiprotonic helium \cite{2006056}.
Figure~\ref{fig:nvsz} gives iso-frequency curves corresponding to the
spacing between adjacent Bohr energy levels ($n$ to $n-1$) in the
2-dimensional parameter space of the principal quantum number $n$ and
the nuclear charge $Z$ for hydrogen-like ions.  Much of this space is
accessible to optical frequency synthesizers based on mode-locked
femtosecond lasers, which readily provide ultra-precise reference rulers
spanning the near-infrared and visible region of the optical spectrum
(530 nm--2100 nm).  Diverse techniques in spectroscopy (such as
double-resonance methods) broaden the range of applications.  Even when
the absolute accuracy is limited by the primary frequency standard (a
few parts in $10^{16}$), optical frequency combs can enable relative
frequency measurements with uncertainties approaching 1 part in
$10^{19}$ over 100 THz of bandwidth \cite{2004013}.  The precisely
controlled pulse train from a femtosecond laser can also be used
directly to probe the global atomic structure, thus integrating the
optical, terahertz, and radio-frequency domains \cite{2004312}.

There are simplifications in the theory of energy levels of Rydberg
states of hydrogen-like ions that, in some cases, allow calculations to
be made at levels of accuracy comparable to these breakthroughs in
optical metrology.  In the following, we write the known theoretical
expressions for the energy levels of these ions, describe and give
results of a calculation that eliminates the largest source of
uncertainty, and list the largest remaining sources of uncertainty.  We
also make numerical predictions for a transition in two different ions
as illustrations, look at the natural line width, and discuss what might
be learned from comparison of theory and experiment. 

In a high-$n$ Rydberg state of a hydrogen-like atom with nuclear charge
$Z$ and angular momentum $l = n-1$, the probability of the electron
being within a short distance $r$ from the origin is of order $(2Zr/n
a_0)^{2n+1}/(2n+1)!$, where $a_0$ is the Bohr radius.  Due to this
strong damping near the origin, effects arising from interactions near
or inside the nucleus are negligible, including the effect of the finite
size of the nucleus.

For $l\ge 2$, the theoretical energy levels can be accurately expressed
as a sum of the Dirac energy with nuclear motion corrections $E_{\rm
DM}$, relativistic recoil corrections $E_{\rm RR}$, and radiative
corrections $E_{\rm QED}$: $E_n = E_{\rm DM} + E_{\rm RR} + E_{\rm
QED}$.  Reviews of the theory and references to original work are given
in \cite{1990020,2001057,2005191}.  The difference between the Dirac
eigenvalue and the electron rest energy is proportional to

\begin{eqnarray}
\alpha^2D &=&
\left[1+\frac{(Z\alpha)^2}{(n-\delta)^2\,}\right]^{-\frac{1}{2}} - 1
\\ \nonumber 
D&=& - \frac{Z^2}{2n^2} +
\left(\frac{3}{8n} - \frac{1}{2j+1}\right)
\frac{Z^4\alpha^2}{n^3} + \dots \, , \quad
\end{eqnarray}
where $\alpha$ is the fine-structure constant, $\delta = |\kappa| -
\sqrt{\kappa^2 - (Z\alpha)^2}\,$, $\kappa = (-1)^{l+j+1/2}(j+1/2)$ is
the Dirac spin-angular quantum number, and $j$ is the total angular
momentum quantum number.  The energy level, taking into account the
leading nuclear motion effects, but not including the electron or
nucleus rest energy, is given by \cite{2001057}
\begin{eqnarray}
E_{\rm DM} &=& 2hcR_\infty\Bigg[\mu_{\rm r}D
-\frac{r_{\rm N}\mu_{\rm r}^3 \alpha^2}{2}D^2
+ \frac{r_{\rm N}^2\mu_{\rm r}^3Z^4 \alpha^2 }{2n^3 \kappa(2l+1)} 
\Bigg] ,\quad
\label{eq:relred}
\end{eqnarray}
where $h$ is the Planck constant, $c$ is the speed of light, $R_\infty =
\alpha^2 m_{\rm e}c/2h$ is the Rydberg constant, $r_{\rm N} = m_{\rm
e}/m_{\rm N}$ is the ratio of the electron mass to the nucleus mass, and
$\mu_r = 1/(1+r_{\rm N})$ is the ratio of the reduced mass to the
electron mass.

Relativistic corrections to Eq.~(\ref{eq:relred}) associated with motion
of the nucleus are classified as relativistic-recoil 
corrections. 
For the states with $l\ge2$ under consideration here, we have
\begin{eqnarray}
E_{\rm RR} &=& 2hcR_\infty\frac{r_{\rm N}Z^5\alpha^3}{ \pi n^3}
 \bigg\{\mu_{\rm r}^3 \bigg[
-\frac{8}{3}\ln k_0(n,l) 
\nonumber \\ && \qquad\qquad\qquad\qquad
-\frac{7}{3l(l+1)(2l+1)} \bigg]
\label{eq:erelrec}
\\  \vbox to 0.7 cm {}&&
+ \pi Z \alpha
\left[3-{l(l+1)\over n^2}\right]
{2\over(4l^2-1)(2l+3)} + \dots \bigg\} \, ,
\nonumber
\end{eqnarray}
where $\ln k_0(n,l)$ is the Bethe logarithm.  We
assume that the uncertainty due to uncalculated higher-order terms is
$Z\alpha\,\ln{(Z\alpha)^{-2}}$ times the contribution of the last term
in Eq.~(\ref{eq:erelrec}).

Quantum electrodynamics (QED) corrections for high-$l$ states 
are summarized as
\begin{eqnarray}
&&E_{\rm QED} = 2hcR_\infty\frac{Z^4\alpha^2}{n^3}
\bigg\{ -\mu_{\rm r}^2
\frac{a_{\rm e}}{\kappa(2l+1)}
\label{eq:qed}
\\ && \quad
+\mu_{\rm r}^3 \,
\frac{\alpha}{\pi}
\bigg[-\frac{4}{3} \ln{k_0(n,l)}
+\frac{32}{3}\,\frac{3n^2-l(l+1)}{n^2}
\nonumber\\ && \quad\nonumber
\times\frac{(2l-2)!}{(2l+3)!}
\,(Z\alpha)^2\ln{\left[\frac{1}{\mu_{\rm r}(Z\alpha)^2}\right]}
+(Z\alpha)^2 \, G(Z\alpha)
\bigg] \bigg\} \, ,
\end{eqnarray}
where $a_{\rm e}$ is the electron magnetic moment anomaly and
$G(Z\alpha)$ is a function that contains higher-order QED corrections.
Equation (\ref{eq:qed}) contains no explicit vacuum polarization
contribution because of the damping of the wavefunction near the origin.
Also in that equation, the uncertainties in the theory of $a_{\rm e}$
may be eliminated by using the experimental value $a_{\rm e} =
1.159\,652\,180\,85(76)\times 10^{-3}$ obtained with a one-electron
quantum cyclotron \cite{2006081}.

The leading terms in $G(Z\alpha)$ are expected to be of the form
\begin{eqnarray}
G(Z\alpha) &=& A_{60} + A_{81}(Z\alpha)^2\ln{(Z\alpha)^{-2}}
+ A_{80}(Z\alpha)^2 + \dots
\nonumber\\
&& + \frac{\alpha}{\pi}\,B_{60}  + \dots
+ \left(\frac{\alpha}{\pi}\right)^2\,C_{60} + \dots
\label{eq:gval}
\end{eqnarray}

The coefficients indicated by the letter $A$ arise from the one-photon
QED corrections; $A_{60}$ and $A_{81}$  arise from the self energy, and
$A_{80}$ arises from both the self energy and the long-range component
of the vacuum polarization.  The term $A_{60}$ has been calculated for
many states with $l \le 8$, but not for higher-$l$ states before this
work.  The uncertainty introduced by this term if it were not
calculated, based on plausible extrapolations from lower-$l$ known
values, would be the largest uncertainty in the theory and larger than
the uncertainty from the Rydberg constant.  The higher-order coefficient
$A_{81}$ and the self-energy component of the coefficient  $A_{80}$ are
not known, but can be expected to be small.  The vacuum polarization
contribution to $A_{80}$ is known \cite{1956001} and is extremely small.
The coefficient $B_{60}$ arises from two-photon diagrams and has not
been calculated for high-$l$ states, but a comparison of calculated
values of $B_{60}$ \cite{2006316} and $A_{60}$ \cite{2003187} for
$l\le5$, suggests it has a magnitude of roughly  $4A_{60}$, which is
used as the associated uncertainty.  Further, a term proportional to
$(\alpha/\pi)B_{61} \,\ln{(Z\alpha)^{-2}}$, that is nonzero for S and P
states, vanishes for higher-$l$ states \cite{2005212}.  The term
$C_{60}$ is expected to be the next three-photon term, in analogy with
the two-photon terms.
 
In order to eliminate the main source of theoretical uncertainty in the
energy levels, we have calculated the value of $A_{60}$ for a number of
Rydberg states.  This calculation uses methods from field theory, {\it
i.e.,} nonrelativistic QED (NRQED) effective operators which facilitate
the calculation \cite{1986022}, and methods from atomic physics to
handle the extensive angular momentum algebra in the higher-order
binding corrections of near-circular Rydberg states.  Distinct
contributions to the self-energy from high- and low-energy virtual
photons, are matched using an intermediate cutoff parameter
\cite{pp07wj}.  For near-circular Rydberg states, the radial wave
functions have at most a few nodes, yet the calculation of $A_{60}$
coefficients for these states is much more involved than for low-lying
states.  The reason is that in using the Sturmian decomposition of the
hydrogen Coulomb Green function, as done for lower-$n$ states, the
radial integrations lead to sums over hypergeometric functions with high
indices, which in turn give rise to an excessive number of terms.  For
states with $n = 8$, there are of order $10^5$ terms in intermediate
steps, which is roughly two orders of magnitude more terms than for
states with $n=2$ \cite{2003108}. This trend continues as $n$ increases
making calculation at high $n$ with this conventional method
intractable.

Here we report that the calculation has been done with a combined
analytic and numerical approach based on lattice methods by using a
formulation of the Schr\"odinger-Coulomb Green function on a numerical
grid \cite{PhRvA.40.5559}. Provided quadruple precision ($\sim$ 32
significant digits) in the Fortran code is used, and provided a large
enough box to represent the Rydberg states on the grid is used, the
positive continuum of states can be accurately represented by a
pseudo-spectrum of states with positive discrete energies. With this
basis set, the virtual photon energy integration can be carried out
analytically for each pseudo-state using Cauchy's theorem.  This solves
the problem of the calculation of the relativistic Bethe logarithms
without the need for the subtraction of many pole terms, which would
otherwise be necessary if the virtual photon energy were used as an
explicit numerical integration variable.  The results of this
calculation for a number of states with $n = 13$ to 16 are given in
Table~\ref{tab:a60}.  \newcommand\str{\hbox to 4.4pt {}} \begin{center}
\begin{table}[t] \caption{Calculated values of the constant $A_{60}$.
The numbers in parentheses are standard uncertainties in the last
figure.} \label{tab:a60} \begin{tabular}{ c @{\str} c  @{\str} | c
@{\str} c @{\str} D{.}{.}{1.9}@{\qquad\quad} | c @{\str} c @{\str}
D{.}{.}{1.11}@{\qquad}} \toprule $n$ & $l$ & $2j$ & $\kappa$ &
\multicolumn{1}{c}{$A_{60}$} & $2j$ & $\kappa$ &
\multicolumn{1}{c}{$A_{60}$} \T\B \\ \colrule 13  & 11  & $21$  & 11 &
0.679\,575(5)\times 10^{-5} & $23$ & -12 & 4.318\,998(5)\times 10^{-5}
\T \\ 13  & 12  & $23$  & 12 &  0.469\,973(5)\times 10^{-5} & $25$ & -13
& 2.729\,475(5)\times 10^{-5}  \\ 14  & 12  & $23$  & 12 &
0.410\,825(5)\times 10^{-5} & $25$ & -13 & 2.979\,937(5)\times 10^{-5}
\\ 14  & 13  & $25$  & 13 &  0.296\,641(5)\times 10^{-5} & $27$ & -14 &
1.945\,279(5)\times 10^{-5}  \\ 15  & 13  & $25$  & 13 &
0.252\,108(5)\times 10^{-5} & $27$ & -14 & 2.116\,050(5)\times 10^{-5}
\\ 15  & 14  & $27$  & 14 &  0.189\,309(5)\times 10^{-5} & $29$ & -15 &
1.420\,631(5)\times 10^{-5}  \\ 16  & 14  & $27$  & 14 &
0.155\,786(5)\times 10^{-5} & $29$ & -15 & 1.540\,181(5)\times 10^{-5}
\\ 16  & 15  & $29$  & 15 &  0.121\,749(5)\times 10^{-5} & $31$ & -16 &
1.059\,674(5)\times 10^{-5} \B \\ \botrule \end{tabular} \end{table}
\end{center}

\vskip -25 pt 
We incorporate the results for $A_{60}$ to numerically
evaluate the theoretical prediction for the frequency of the transition
between the state with $n=14$, $l=13$, $j=\frac{27}{2}$ and the state
with $n=15$, $l=14$, $j=\frac{29}{2}$ in the hydrogen-like ions He$^+$
and Ne$^{9+}$.  The constants used in the evaluation are the 2006 CODATA
recommended values \cite{constants}, with the exception of the neon
nucleus mass $m({\rm ^{20}Ne^{10+}})$ which is taken from the neon
atomic mass \cite{2003253}, corrected for the mass of the electrons and
their binding energies.  Values of the various contributions and the
total are given as frequencies in Table~\ref{tab:fhene}.  Standard
uncertainties are listed with the numbers where they are non-negligible.
The theory is sufficiently accurate that the largest uncertainty arises
from the Rydberg frequency $cR_\infty$, which is a factor in all of the
contributions.  There is no uncertainty from the Planck constant, since
$\nu = (E_{15}-E_{14})/h$.  

Table~\ref{tab:uncs} gives sources and estimates of the various known
uncertainties in the theory.  To put them in perspective, in hydrogen,
the relative uncertainty from the two-photon term $B_{60}$ for the
1S--2S transition is of the order of $10^{-12}$ due to disagreement
between different calculations, whereas in the $n=14$ to $n=15$
Rydberg transition it is likely to be roughly $5\times10^{-19}$, based
on the smallness of the calculated value of the $A_{60}$ coefficient.
The improved convergence of the expansion of the QED corrections in
powers of $Z\alpha$ is indicated by the fact that $A_{60}$ is smaller by
a factor of about $10^{6}$ for the Rydberg states than the value
$A_{60}\approx-30$ for S states.

\newcommand\strr{\hbox to 15pt {}}
\begin{center} \begin{table}[t] 
\caption{Transition frequencies between the highest-$j$ states 
with $n=14$ and $n=15$ in hydrogen-like helium and hydrogen-like
neon.} \label{tab:fhene} 
\begin{tabular}{ l @{\strr}D{.}{.}{5.15} @{\strr} D{.}{.}{5.15} } 
\toprule 
Term & \multicolumn{1}{c}{$^4$He$^+$~$\nu$(THz)} & 
\multicolumn{1}{c}{$^{20}$Ne$^{9+}$~$\nu$(THz)} \vb \\ 
\colrule 
$E_{\rm DM}$  &  8.652\,370\,766\,008(58)  &  216.335\,625\,5746(14) \vb \\ 
$E_{\rm RR}$  &  0.000\,000\,000\,000  &  0.000\,000\,000\,1     \\ 
$E_{\rm QED}$ &  -0.000\,000\,001\,894 &  -0.000\,001\,184\,1    \\ 
Total         &  8.652\,370\,764\,114(58) &  216.335\,624\,3907(14)    \\ 
\botrule 
\end{tabular} \end{table} \end{center}

\begin{center} \begin{table}[b] 
\caption{Sources and estimated relative standard uncertainties
in the theoretical value of the transition
frequency between the highest-$j$ states with $n=14$ and
$n=15$ in hydrogen-like helium and hydrogen-like neon.} \label{tab:uncs} 
\begin{tabular}{ l @{\qquad\qquad}D{.}{.}{4} @{\qquad} c @{\qquad}
D{.}{.}{4}@{\qquad} } 
\toprule 
Source & \multicolumn{1}{c}{He$^+$} && 
\multicolumn{1}{c}{Ne$^{9+}$}\vb \\ 
\colrule 
Rydberg constant & 6.6\times10^{-12} && 6.6\times10^{-12} \vb \\ 
Fine-structure constant & 7.0\times10^{-16} && 1.7\times10^{-14} \\ 
Electron-nucleus mass ratio& 5.8\times10^{-14} && 1.2\times10^{-14} \\ 
$a_{\rm e}$ & 1.4\times10^{-19} && 3.5\times10^{-18} \\ 
Theory: $E_{\rm RR}$ higher order & 6.2\times10^{-17} && 2.4\times10^{-14} \\ 
Theory: $E_{\rm QED}~A_{81}$ & 1.7\times10^{-18} && 1.6\times10^{-14} \\ 
Theory: $E_{\rm QED}~B_{60}$ & 8.6\times10^{-18} && 5.4\times10^{-15} \\ 
\botrule 
\end{tabular} \end{table} \end{center}

\vskip -50 pt 
The QED level shift given by Eq.~(\ref{eq:qed}) is
understood to be the real part of the radiative correction, while the
complete radiative correction to the level ${\cal E}_{\rm QED} = E_{\rm
QED} - \ri\,\iG/2$ is complex and includes an imaginary part
proportional to the rate $A = \iG/\hbar$ for spontaneous radiative decay
of the level to all lower levels.  For the highest-$l$ state with
principal quantum number $n$, the dominant decay mode is an E1 decay to
the highest-$l$ state with principal quantum number $n-1$ \cite{BS}.
Formulas in Ref.~\cite{BS} give the nonrelativistic expression for the
decay rate, which can also be derived from the nonrelativistic limit of
the imaginary part of the level shift \cite{2005140}.

\begin{figure}[t]
\resizebox{!}{2.3 in}{ \includegraphics{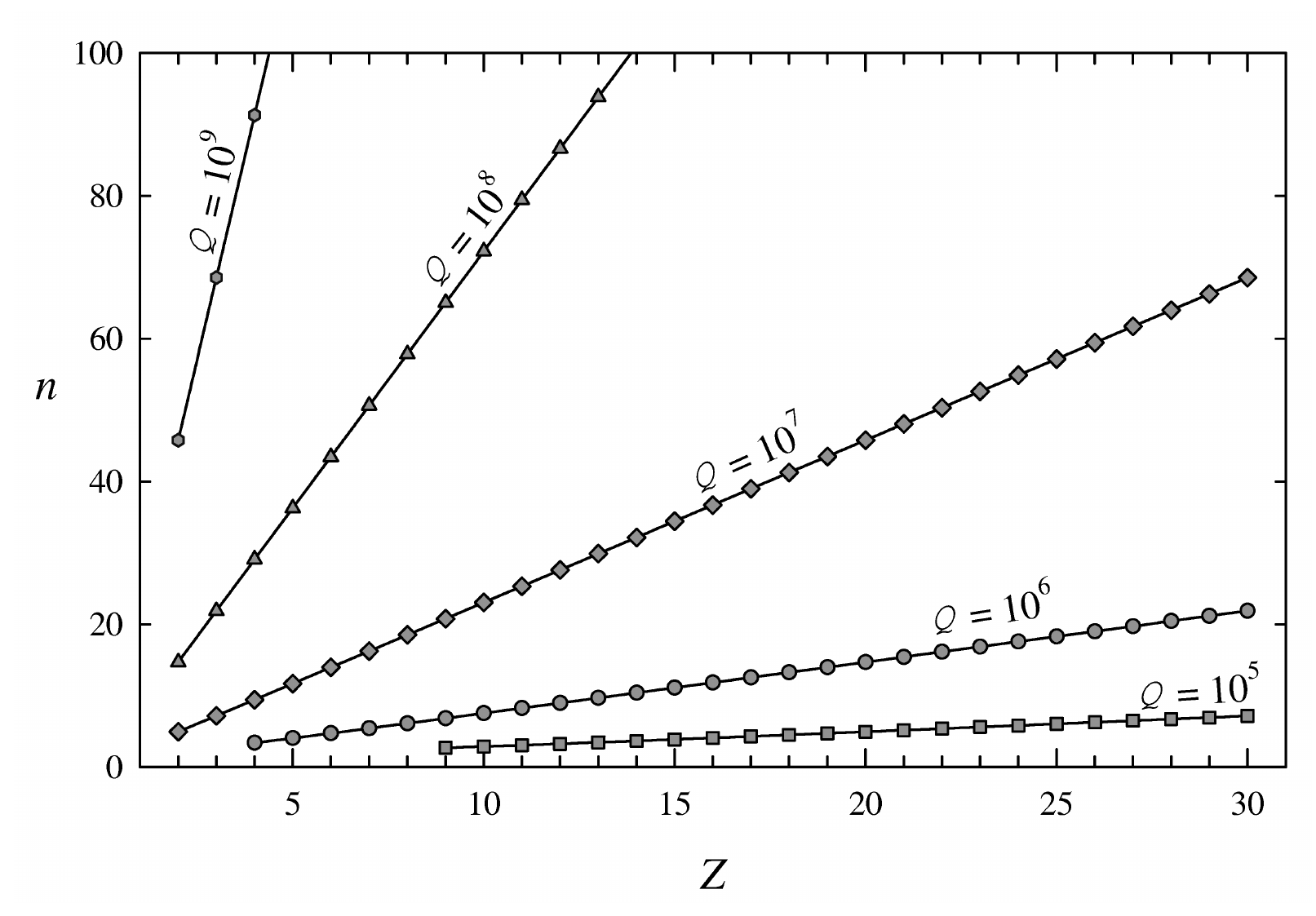}}
\caption{ \label{fig:widths}
Graph showing values of $Z$ and approximate $n$ that give a specified
ratio of the transition frequency to the natural width of the transition
resonance between circular states with principal quantum number $n$ and
$n-1$ in a hydrogen-like ion with nuclear charge $Z$.}
\end{figure}

As a first approximation, for transitions between states with quantum
numbers $n$ and $n-1$ the ratio of the transition energy to the width of
the line, is given by 
\begin{eqnarray}
Q &=& \frac{E_n -
E_{n-1}}{\iG_n+\iG_{n-1}} \rightarrow \frac{3n^2}{4\alpha(Z\alpha)^2} +
\dots \, , 
\label{eq:asym} 
\end{eqnarray} 
where the expression on the right is the asymptotic form as
$n\rightarrow \infty$ of the nonrelativistic value.
Figure~\ref{fig:widths} gives a contour plot of the values of $n$ and
$Z$ that give a specified value of $Q$ based on the nonrelativistic
result in Ref.~\cite{BS}.  This is just a rough indication, since
transitions with smaller $l$ values will generally have a smaller $Q$,
whereas transitions with a change of $n$ greater than 1 will have a
larger $Q$.  The effect of possible asymmetries of the line shape on the
apparent resonance center has been shown to be small by Low
\cite{1952012}, of order $\alpha(Z\alpha)^2\,E_{\rm QED}$.  For the
1S--2S transition in hydrogen, such effects are indeed completely
negligible at the current level of experimental accuracy \cite{2002211}.
However, for Rydberg states of hydrogen-like ions, particularly at
higher-$Z$, asymmetries in the line shape, some of which depend on
details of the experiment, may be significant, and can be calculated if
necessary.

Recent advances in atomic-molecular-optical physics have generated an
array of tools and techniques useful for engineering highly simplified
atomic systems \cite{AMO2010}.   In particular, observations of cold
antihydrogen production at CERN illustrate two ways for a cooled
ion/antiproton to capture an electron/positron in high-$l$ Rydberg
states, either by three-body recombination or by charge exchange
\cite{2005236}.  Properties of atomic cores have also been studied using
a double-resonance detection technique to observe the fine structure of
Rydberg states produced by charge exchange in a fast beam of
highly-charged ions \cite{PhRvA.75c2523}.  Using electron cooling
\cite{2005236} (and charge exchange), cold hydrogen-like ions can be
recombined in high-$l$ Rydberg states from a variety of bare ions
extracted from sources such as an electron beam ion source/trap
(EBIS/T).  Although two-photon spectroscopy is possible in certain
cases, if the ions are confined in a trap within a region smaller than
about half the wavelength of the radiation exciting the transition,
Dicke narrowing also eliminates the first-order Doppler shift
\cite{itano1995}.  Assuming $T = 100$ K, the relative second-order
Doppler shift is about $3.5\times 10^{-12}$ for He$^+$ and
$7\times10^{-13}$ for Ne$^{9+}$.  Temperatures in the range 4 K $< T <
77$ K are obtainable in cryogenic ion traps by resistive cooling
\cite{itano1995} and by electron or positron cooling \cite{2005236}.
For lower temperatures ($T < 1$ K), sympathetic laser cooling methods
can be used \cite{itano1995}.

Of the variety of $(n,l,Z)$ combinations of hydrogen-like ions, circular
Rydberg states of low-$Z$ ions seem the most favorable for a comb-based
determination of the Rydberg constant.  On the other hand, some
perturbations are smaller and line-widths are larger in heavier ions.
Hence, using ions with a variety of $(n,Z)$ combinations could be useful
for experimental optimization and consistency checks, as well as for
extending diversity of experiments used to determine fundamental
constants and test theory.

UDJ acknowledges support from the Deutsche Forschungsgemeinschaft
(Heisenberg program).  Mr. Frank Bellamy provided assistance with some
of the numerical evaluations.

\end{document}